\documentclass[twocolumn,showpacs,aps,prl,superscriptaddress]{revtex4}

\usepackage{graphicx}
\usepackage{amsmath}

\begin{document}

\title{Observation of Inter-Valley Gap Anomaly of Two Dimensional electrons in Silicon}

\author{K. Lai}
\affiliation{Department of Electrical Engineering, Princeton
University, Princeton, New Jersey 08544}
\author{W. Pan}
\affiliation{Sandia National Laboratories, Albuquerque, NM
87185}
\author{D.C. Tsui} \affiliation{Department of Electrical
Engineering, Princeton University, Princeton, New Jersey 08544}
\author{S. Lyon}
\affiliation{Department of Electrical Engineering, Princeton
University, Princeton, New Jersey 08544}
\author{M. M\"uhlberger}
\affiliation{Institut f\"ur Halbleiterphysik, Universit\"at Linz, A-4040 Linz, Austria}
\author{F. Sch\"affler}
\affiliation{Institut f\"ur Halbleiterphysik, Universit\"at Linz, A-4040 Linz, Austria}

\date{\today}

\begin{abstract}

We report here a systematic study of the energy gaps at the odd-integer quantum Hall states $\nu=3$ and 5 under tilted magnetic (B) fields in a high quality Si two-dimensional electron system. Out of the coincidence region, the valley splitting is independent of the in-plane B fields. However, the $\nu=3$ valley gap differs by about a factor of 3 ($\Delta_v\sim$ 0.4K vs. 1.2K) at either side of the coincidence. More surprisingly, instead of reducing to zero, the energy gaps at $\nu=3$ and 5 rise rapidly when approaching the coincidence angles. We believe that such anomaly is related to strong coupling of the nearly degenerate Landau levels.

\end{abstract}
\maketitle

There has been a revival of research interest in understanding the valley degree of freedom in silicon (Si) since the original proposal of utilizing spins in Si as quantum bits\cite{kane,koiller,boykin,xiao}. In bulk Si, electrons occupy six degenerate valleys in the Brillouin zone. For the two-dimensional electron system (2DES) confined in a (001) Si quantum well\cite{schaffler}, this six-fold degeneracy is lifted, with the two out-of-plane valleys (the $\pm$k$_z$ valleys) lower in energy than the four in-plane valleys. Further lifting of the remaining two-fold degeneracy has long been noticed, while its physical origin remains controversial\cite{ohkawa,sham,ando}. In light of the quantum computation, a better understanding of this valley splitting now becomes more important since the existence of these two nearly degenerate $\pm$k$_z$ valleys has been considered a potential source of spin decoherence\cite{koiller}.

Previous experimental investigations on the valley splitting were largely performed in Si metal-oxide-semiconductor field-effect transistors (MOSFETs), where the disorder effect is strong\cite{nicholas,pudalov,khrapai}. In the past decade, significant progress has been made on high quality modulation doped Si/SiGe heterostructures\cite{schaffler}, and it has become possible to study the valley splitting with much reduced disorder\cite{weitz,koester,dunford,shlimak,goswami,wilde}. In this letter, we report experimental results of the energy gap measurement under tilted magnetic (B) fields in a high mobility Si 2DES. Consistent with previous observations by Weitz $et$ $al$\cite{weitz}, we observed that the valley splitting is independent of in-plane B field out of the so-called coincidence region where two Landau levels cross each other at the Fermi level. However, the $\nu=3$ valley gap differs by about a factor of 3 ($\Delta_v\sim$ 0.4K vs. 1.2K) at either side of the coincidence. Even more surprisingly, in the coincidence region the inter-valley gaps at $\nu=3$ and 5, instead of reducing to zero as expected in an independent-electron model, rise rapidly when approaching the coincidence angles. We believe that such anomaly is related to strong coupling of the Landau levels close in energy in the coincidence region.

The specimen in our experiment is a modulation doped Si quantum well sandwiched between Si$_{0.75}$Ge$_{0.25}$ barriers. Details on the sample structure and fabrication process can be found in Ref. \cite{lai1}. When cooled down to T=50mK in dark, the sample has an electron density n=1.4$\times$10$^{11}$cm$^{-2}$ and mobility $\mu$=1.9$\times$10$^5$cm$^2$/Vs (labeled as sample S1 in the following text). After a short period of low temperature illumination, the density and mobility increase to 2.4$\times$10$^{11}$cm$^{-2}$ and 2.5$\times$10$^5$cm$^2$/Vs, respectively (labeled as sample S2, thereafter). Magnetotransport measurements were carried out in a dilution refrigerator equipped with an in-situ rotating stage. The typical temperature range for thermal activation measurements is from 50mK to about 500mK.

Fig. 1 shows the diagonal resistivity $\rho_{xx}$ of sample S1 in a perpendicular magnetic field as a function of the Landau level (LL) filling factor $\nu$, at T=55mK. For $\nu>16$, only minima corresponding to $\nu$=4(N+1) are resolved in the Shubnikov-de Haas oscillations, where N is the LL index. The integer quantum Hall effect (IQHE) states with $\nu$=4N+2 develop for N$<$4 and the odd-integer states $\nu=3$ and 5 show up in the last two LLs. Such a complex IQHE series can be understood by comparing the energy scales in Si 2DES, the cyclotron gap E$_C$, the Zeeman gap E$_Z$ and the inter-valley gap $\Delta_v$. As illustrated schematically in the inset of Fig. 1, in a perpendicular B field, different LLs are separated by the cyclotron gap $E_C=\hbar\omega_C=\hbar eB/m^{\ast}$, where m$^{\ast}$ is the effective mass. For each LL, the Zeeman coupling of electron spins to the B field opens up a spin gap E$_Z$ = g$^{\ast}\mu_B$B, where g$^{\ast}$ is the effective g-factor and $\mu_B$ the Bohr magneton, between the spin-up and down electrons. Each spin level is further split by a small valley gap $\Delta_v$. In general, E$_C>E_Z>\Delta_v$.

\begin{figure}[!t]
\begin{center}
\includegraphics[width=3.2in,trim=0.3in 0.5in 0.2in 0.2in]{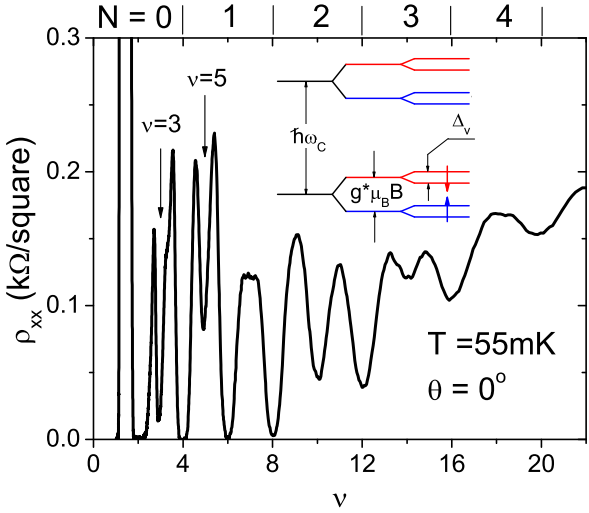}
\end{center}
\caption{\label{1}Longitudinal resistivity $\rho_{xx}$ for S1 in a perpendicular magnetic (B) field, taken at 55mK. Two odd integer Quantum Hall states, which we studied in great detail in this letter, are indicated by arrows. The bottom x-axis is the Landau level (LL) filling factor $\nu$ . The LL index is marked on the top.}
\end{figure}

In the tilted B field measurement, the normal of the 2D plane is tilted from the field direction by an angle, $\theta$. In Fig. 2a, $\rho_{xx}$ around $\nu=3$ is shown at a few selected angles for sample S1. The x-axis is the perpendicular magnetic field, B$_\perp$, and traces under tilted B field are shifted vertically for clarity. At $\theta=0^o$, $\nu=3$ is a well-developed quantum Hall state, manifested by a deep $\rho_{xx}$ minimum and a quantized Hall plateau (not shown). In fact, in our experiment, the $\nu=3$ state remains as a good QHE state at all the tilt angles. Starting from $\theta\sim64.1^o$, a hysteretic resistance spike develops around the $\nu=3$ minimum. It moves from the lower B side to the higher B side of the $\rho_{xx}$ minimum as the tilt angle increases, and eventually becomes a sharp shoulder when $\theta>65.0^o$. This resistance spike has been observed before and is a signature of the first order magnetic transition between quantum Hall ferromagnets\cite{eom,de,jungwirth}. We believe that this mechanism is also responsible for the appearance of the resistance spike in our system. Details on the resistance spike in our Si sample are given elsewhere\cite{lai2}.

\begin{figure}[!t]
\begin{center}
\includegraphics[width=3.2in,trim=0.3in 0.5in 0.2in 0.2in]{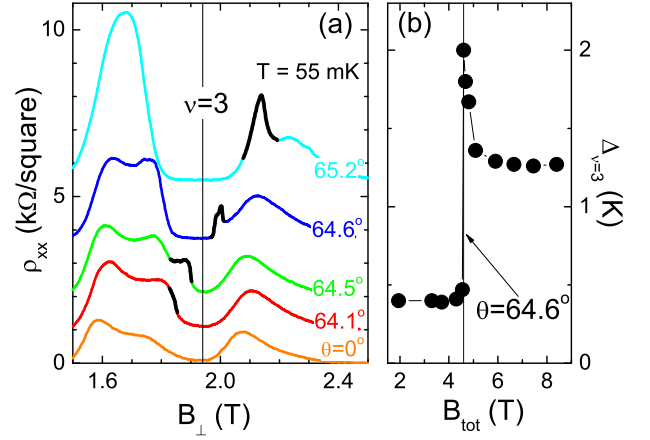}
\end{center}
\caption{\label{2} {\bf a}. (sample S1) Selected magnetotransport traces as a function of B$_\perp$ near the coincidence point. Traces are shifted vertically for clarity. A resistance spike is clearly seen around the $\nu=3$ minimum, highlighted by the thick black lines. It moves from the lower B-field to high B-field side of $\rho_{xx}$ the minimum as $\theta$ increases. {\bf b}. Activation energy gap at $\nu=3$. It is nearly constant before and after $\theta$=64.6$^o$, where a sudden jump occurs. Note that the gap within the jump was not measured due to the appearance of the resistance spike.}
\end{figure}

The most remarkable feature in Fig. 2a is the sudden change in the strength of the $\nu=3$ state. For an angular change of merely 0.1$^o$, $\rho_{xx}$ at $\nu=3$ evolves from a minimum at 64.5$^o$ to a wide "zero-resistance" state at 64.6$^o$, indicating an abrupt strengthening of this IQHE state. To quantify this observation, we have measured its energy gap $\Delta_3$, deduced from the temperature dependence of the diagonal resistivity $\rho_{xx}$$\propto$ exp(-$\Delta_3$/2k$_B$T). During the temperature dependence measurement, extra care was taken to make sure that at each point the spike, if appeared, was always at one side of $\nu=3$. In Fig. 2b, $\nu=3$ is plotted as a function of the total field B$_{tot}$. It is nearly constant, $\sim$0.4K, at small B$_{tot}$. At $\theta=64.6^o$, it jumps to 2.0K. On the higher B$_{tot}$ side of $\theta=64.6^o$, a qualitatively similar B dependence is also observed for $\Delta_3$. It is $\sim$1.25K at large B$_{tot}$, and increases rapidly when approaching $\theta=64.6^o$. Interestingly, the B$_{tot}$ dependence of $\Delta_3$ appears to be highly asymmetric: On the left-hand side of coincidence, $\Delta_3$ increases from 0.47K to 2K in 0.05T (or a rate of $\sim$31K/T), while on the right-hand side, it increases from 1.36K to 2K in 0.48T (or a rate of $\sim$1.3K/T).

This angular dependence of $\Delta_3$ in the coincidence region is unexpected and anomalous. To understand this anomaly, we start from the non-interacting single particle model. For a constant B$_{\perp}$, B$_{tot}$ increases as B$_{tot}$ = B$_\perp$/cos$\theta$. Consequently, the ratio of the Zeeman splitting E$_Z$ = g$^\ast$$\mu_B$B$_{tot}$ to the Landau level separation E$_C$ = $\hbar\omega_C$ = $\hbar$eB$_\perp$/m$^\ast$ increases with increasing $\theta$. A coincidence happens when two energy levels from different LLs cross each other. In Fig. 3b, a Landau fan diagram is constructed and each level evolves as E = (N+1/2)$\hbar\omega_C\pm$1/2$\Delta_v\pm$1/2g$^\ast$$\mu_B$$B_{tot}$. For sample S1, the $\nu=3$ state occurs at B$_\perp$=2T and therefore $\hbar\omega_C$ = 13.4K, using the Si band mass m$^\ast$=0.2m$_e$. Two valley splittings, $\Delta_v$(N=0,$\downarrow$) and $\Delta_v$(N=1,$\uparrow$), (see Fig. 3) are involved in the angular dependence of $\nu=3$. To determine $\Delta_v$(N=0,$\downarrow$), we note that at $\theta$=0$^o$ the $\nu=3$ energy gap is presumably due to the valley-flip between the + and -- valleys \cite{valley}. Thus, by neglecting the level broadening, $\Delta_v$(N=0,$\downarrow$) is equal to $\Delta_3$ at $\theta$=0$^o$. In the tilted LL diagram, the two valley-split states are parallel to each other since the valley splitting is shown by Weitz $et$ $al$ \cite{weitz} to depend only on B$_\perp$. To determine $\Delta_v$(N=1,$\uparrow$), we note that it is equal to $\Delta_3$ at high tilt angles, after the coincidence. Thus, $\Delta_v$(N=1,$\uparrow$)=1.25K is chosen. This assignment is consistent with the energy gap measurement at $\nu=5$ in S2. There, at B$_\perp$=2T, the inter-valley gap for (N=1,$\uparrow$), is also $\sim$1.25K. Finally, to finish the construction of the LL diagram, we need to know g$^\ast$. Several techniques have been explored in determining g$^\ast$ and here we use the coincidence method, i.e., using the condition of E(N=0,$\downarrow$,+) = E(N=1,$\uparrow$,+) to deduce g$^\ast$=[$\hbar\omega_C$-1/2$\Delta_v$(N=1,$\uparrow$)+1/2$\Delta_v$(N=0,$\downarrow$)]/$\mu_BB_{tot}$. This technique requires the identification of $\theta_C$ or B$_{tot}$. Usually, we rely on the experimental signature that $\rho_{xx}$ becomes a local maximum, since at the coincidence the energy gap vanishes. However, in our sample, the $\nu=3$ state is a good quantum Hall state at all angles and its minimum never disappears. Consequently, we turn to the energy gap measurement. It is observed that $\Delta_3$ increases rapidly from both sides when $\theta$ approaches 64.6$^o$, which we take as the coincidence angle. After identifying this coincidence angle, g$^\ast$ = 4.1 is then deduced. This value is consistent with previous reports on Si/SiGe heterostructures\cite{weitz,koester,dunford,wilde}. The final Landau fan diagram is plotted in Fig. 3b. Three crossing angles, $\theta_1$ and $\theta_2$ corresponding to the level crossing away from the Fermi level and $\theta_C$ corresponding to crossing at the Fermi level, are labeled. From this plot, we can deduce the expected single-particle $\nu=3$ gap, plotted as the gray line in Fig. 3a. For $\theta<\theta_1$ and $\theta>\theta_2$ the energy gap is determined by a valley flip of the same LL and independent of tilt angles. Between $\theta_1<\theta<\theta_2$, the energy gap is now associated with a spin index flip between two different LLs, and this spin-flip gap is expected to become narrowed as $\theta\to\theta_C$ and eventually collapses at $\theta_C$. Remarkably, such an independent-electron picture reproduces almost perfectly our experimental observations for $\theta<\theta_1$ and $\theta>\theta_2$. However, between $\theta_1<\theta<\theta_2$, instead of approaching zero at  $\theta_C$, it rises rapidly. We emphasize here that this gap anomaly occurs right in the region where, under the non-interacting single particle model, the gap is associated with a spin flip.

\begin{figure}[!t]
\begin{center}
\includegraphics[width=3in,trim=0in 0.5in 0in 0.2in]{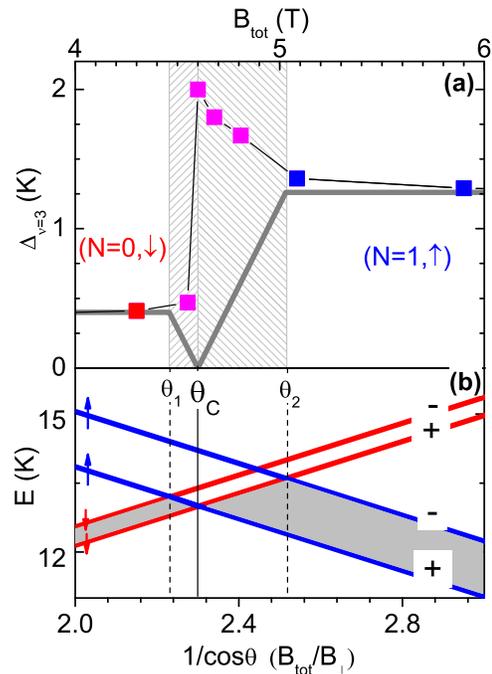}
\end{center}
\caption{\label{3} {\bf a}. The measured $\Delta_3$ gap (squares), close to the coincidence region. The expected single particle gap calculated from Landau fan diagram in Fig. 3b is also shown (gray line) for comparison. {\bf b}. Landau fan diagram for $\nu=3$ in tilted B field. The x-axis is 1/cos$\theta$ or B$_{tot}$/B$_\perp$. The energy levels are labeled with the LL index and spin orientations. The expected single-particle $\nu=3$ gap is shaded for clarity. The three crossing angles relevant to the $\nu=3$ gap are labeled as $\theta_1$, $\theta_C$ and $\theta_2$, respectively. $\theta_1$ and $\theta_2$ are the coincidence angles where level crossing occurs away from the Fermi energy. $\theta_C$ is the usual coincidence angle where level crossing occurs at the Fermi energy.}
\end{figure}

In order to further investigate this inter-valley gap anomaly, we turn to the next odd-integer state $\nu=5$ in sample S2. Again, the inter-valley gap anomaly is observed, as shown in Fig. 4. Out of the coincidence region, the $\nu=5$ gap, $\Delta_5$, is constant, and assumes the values of valley splitting. In the coincidence region, $\Delta_5$ rises rapidly towards the coincidence angles. The two peaks correspond to the first and second coincidences for $\nu=5$. The Landau fan diagram (not shown) was also constructed, and the expected single particle gap for $\nu=5$ is shown as the gray line in Fig. 4. It is clear that similar to $\nu=3$ the gap anomaly occurs right in the coincidence region where the gap is supposed to involve a spin flip. Also, like $\nu=3$, the B$_{tot}$ dependence of $\Delta_5$ close to the coincidence points is asymmetric.

\begin{figure}[!t]
\begin{center}
\includegraphics[width=3.2in,trim=0.3in 0.5in 0.2in 0.2in]{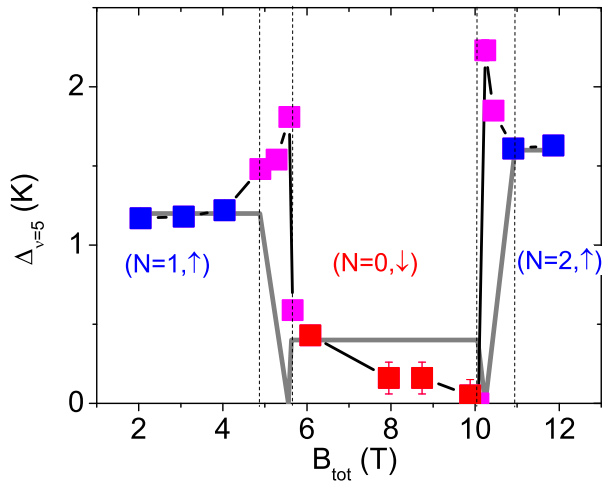}
\end{center}
\caption{\label{4} (sample S2) The B$_{tot}$ dependence of the energy gap at $\nu=5$. The LL and spin indices are indicated for three different regions. Two peaks are observed, corresponding to the first and second coincidences. The expected nominal (gray line) $\nu=5$ gap, calculated according to the independent-electron model, is also shown for comparison. Dotted lines show the boundaries where the gap anomaly occurs.}
\end{figure}

This inter-valley gap anomaly is puzzling. To speculate on its physical origin, we first consider a simple extension of the independent-electron model: The anti-crossing mechanism. Under this picture, when two LLs cross each other, the coupling between these two LLs would open a finite gap at the coincidence point. Usually, this finite gap is small and the evolution of gap vs. B$_{tot}$ is smooth, which is in sharp contrast to our observations that the $\nu=3$ and 5 gaps rapidly increase at the degeneracy points. On the other hand, in our system, a total of four energy levels with different Landau level index, spin, and valley, are brought close in energy near the coincidence by the in-plane magnetic field. Coupling of all four levels may give rise to much more complex off-diagonal terms in the Hamiltonian. If these off-diagonal terms are larger than the diagonal terms, the energy gap may overshoot near the coincidence point and display an asymmetric B$_{tot}$ dependence \cite{kun}. Secondly, since the resistance spike was observed around $\nu=3$, it is natural to speculate that the inter-valley gap anomaly is due to the formation of a quantum Hall ferromagnet (QHF). We notice that under the framework of QHF similar level crossing has been studied in other 2D systems\cite{daneshvar,piazza,muraki}. Among them, the system studied by Muraki $et$ $al$ \cite{muraki} -- a strongly coupled double quantum well system -- closely resembles ours. They also measured the energy gap at $\nu=3$. However, a smooth variation of $\Delta_3$ was observed. Furthermore, the energy gap decreases as the 2DES approaches the coincidence point and saturates at a finite value, different from our observations. By comparing these two experiments carefully, we notice that in the coincidence region $\Delta_3$ in our measurement involves a spin-index flip excitation. On the other hand, in Muraki $et$ $al$'s experiment, the spin index was unchanged. The spin flip may be important in causing the different B field dependent behaviors. In this regard, the inter-valley gap anomaly may originate from the formation of a new spin-related ground state. Perhaps, in the coincidence region, electrons with opposite spins and/or valleys can pair up, condense in a superposition of (N=0,$\downarrow$,+), (N=0,$\downarrow$,--), (N=1,$\uparrow$,+), and (N=1,$\uparrow$,--) -- the four LL levels close in energy, and may even form a superconducting state\cite{cohen,rasolt}. In light of this, the maximal gap of $\sim$ 2.0$\pm$0.2K for both $\nu=3$ (S1) and 5 (S2) can be taken as a measure of the pairing strength. Of course, further theoretical and experimental efforts are needed to test this hypothesis.

In summary, we have carried out a detailed study of the energy gaps at $\nu=3$ and 5 under tilted magnetic fields in a high quality Si 2DES. We observe that the inter-valley gap is nearly constant out of the coincidence region, which can be well understood under an independent-electron picture. Surprisingly, in the coincidence region, the inter-valley gaps at $\nu=3$ and 5 rise rapidly when approaching the coincidence angles. We argue that such anomaly is related to strong coupling of the Landau levels close in energy in the coincidence region.

This work is supported by the NSF and the DOE. Sandia National Labs is operated by Sandia Corporation, a Lockheed Martin Company, for the DOE. Part of the measurements was carried out at the NHMFL. We thank Y. Chen, D. Novikov, R. Bhatt, S. Sondhi, and K. Yang for illuminating discussions.


\begin{thebibliography}{90}
\bibitem{valley} The valley splitting is the energy difference between the two linear combination, labeled as + and - in the text, of the valley wave functions. Here we assume the "+" state is the lower energy configuration.
\end{thebibliography}
\end{document}